# Query Performance Optimization in XML Data Warehouses

**Hadj Mahboubi and Jérôme Darmont**
Université de Lyon (ERIC Lyon 2)
5 avenue Pierre Mendès-France
69767 Bron Cedex
France
hadj.mahboubi@univ-lyon2.fr, jerome.darmont@univ-lyon2.fr

## ABSTRACT

XML data warehouses form an interesting basis for decision-support applications that exploit complex data. However, native-XML database management systems (DBMSs) currently bear limited performances and it is necessary to research for ways to optimize them. In this chapter, we present two such techniques. First, we propose a join index that is specifically adapted to the multidimensional architecture of XML warehouses. It eliminates join operations while preserving the information contained in the original warehouse. Second, we present a strategy for selecting XML materialized views by clustering the query workload. To validate these proposals, we measure the response time of a set of decision-support XQueries over an XML data warehouse, with and without using our optimization techniques. Our experimental results demonstrate their efficiency, even when queries are complex and data are voluminous.

## INTRODUCTION

Decision-support applications aim at facilitating the decision-making process. They collect data from operational databases and various sources, transform them into information available to decision-makers in a consolidated and consistent manner (Kimball & Ross, 2002).

Furthermore, the development of the Web 2.0 and the proliferation of multimedia documents contributed to the analysis of data are not only numerical nor symbolic. Indeed, such data can be represented in various formats (databases, texts, images, sounds, videos...); diversely structured (relational databases, XML document repositories...); originating from several different sources (distributed databases, the Web...); described through several channels or points of view (x-ray photographs and audio diagnosis of a physician, data expressed in different scales or languages...); changing in terms of definition or value (temporal databases, periodical surveys...). We term data that fall in several of the above categories *complex data* (Darmont *et al.*, 2005).

In this context, XML proves a very interesting tool for integrating and warehousing complex data for analysis thanks to its self-description (akin to warehouse metadata) and extensibility features (Darmont et al., 2003). Moreover, XML has become a standard for representing complex business data (Beyer et al., 2005). Hence, many efforts toward XML data warehousing have been achieved in the past few years (Park et al., 2009; Pokorný, 2002).

However, decision-support queries are generally complex because they involve several join and aggregation operations, while most XML-native database management systems (DBMSs) present relatively poor performances when data volume is very large and/or queries are complex.

In classical (i.e., relational) data warehouses, these issues are customarily addressed by indexing data and materializing views (Gupta & Mumick, 2005). Indexes and materialized views are physical data structures that improve data access time. An index allows direct (vs. sequential) access to data, while a materialized view precomputes query results and avoids accessing the whole original data. Both these physical data structures require additional storage space and induce some refreshing process overhead. It is thus crucial to select them wisely.

Several solutions have been proposed for XML data indexing in the literature. However, the existing techniques support single-labeled path expressions within one single XML document (Goldman & Widom, 1997; Chung et al., 2002). Such path expressions help explore an XML document and extract a specific node (element) or sub-tree (subdocument). They cannot perform join operations over several XML documents. In the context of XML data warehouses, decision-support queries are complex and involve several path expressions. Data are also generally distributed into several XML documents due to their large volume. Hence, XML queries should use specific indexes to access these documents.

In the context of relational data warehouses, several studies address the materialized view selection problem (Agrawal et al., 2000; Aouiche et al., 2006). Views that are relevant to materialize are selected to minimize the processing time of a given workload under maintenance cost and/or storage space constraints (Kotidis & Roussopoulos, 1999). Unfortunately, no such view materialization approach exists for XML databases and XML data warehouses in particular.

In this chapter, we propose a new index structure that is specifically adapted to multidimensional XML data warehouses. This structure is able to maintain a star schema[1] of several XML documents and to preserve the information contained in these documents. It is actually a join index that ensures faster execution of decision-support XQueries by eliminating join costs.

Our second contribution consists in adapting Aouiche et al.'s (2006) query clustering-based relational view selection approach to the XML context. We cluster queries and build candidate XML views that can resolve multiple similar queries belonging to the same cluster. Our approach exploits XML-specific cost models to select XML views that are pertinent to materialize.

The remainder of this chapter is organized as follows. We first discuss previous research related to XML indexing and materialized view selection, respectively. Then, we introduce the technical context of our studies, namely the XML data warehouse model we use, before detailing our join index for XML data warehouses and our XML materialized view selection strategy. To validate our proposals, we also present some experimental results. Finally, we conclude this chapter and hint at future research issues.

---

[1] A *star schema* is the simplest data warehouse schema. It consists of a single, central *fact table* linked to peripheral *dimensions*, an analyzed fact thus being described by a combination of dimensions (or analysis axes).

## RELATED WORK

In this section, we discuss the state-of-art XML related to index and materialized view selection approaches.

### XML data indexes

In the following, we assume that an XML document is defined as a labeled graph whose nodes represent document elements or attributes, and edges represent element-subelement (or parent-child) relationships. Edges are labeled with element or attribute names.

Several studies address the issue of XML data indexing (Goldman & Widom, 1997; Milo & Suciu, 1999; Cooper et al., 2001; Chung et al., 2002). They are more particularly devoted to optimize XML path expressions. Generally, they help traverse XML document hierarchies by referencing structural information about these documents. These techniques extract structural information directly from data and create a structural summary that is a labeled, directed graph. Graph schemas can be used as indexes for path queries. In practice, an XML index is a new XML document that is accessed instead of the original document.

Dataguide is a summary structure for semi-structured and XML data (Goldman & Widom, 1997). Its structure describes by one single label all the nodes (elements) whose labels (names) are identical. Its definition is based on targeted path sets, i.e., the set of nodes that are reached by traversing a given path. 1-index clusters nodes that share the same path in the XML data graph (Milo & Suciu, 1999). This process is performed through a bi-similarity relationship. To select labels or express path expressions, a hash table or a B-tree structure is used to index graph labels.

Dataguide and 1-index code all the paths from the root node. Hence, their size may become larger than the original XML document when XML data are represented as graphs (cyclic XML document), which dramatically degrades query performance. A(k)-index, a variant of 1-index, addresses this issue (Kaushik et al., 2002). It is based on the notion of k-bisimilarity[2] and builds an approximate index that reduces index graph size. An A(k)-index can retrieve, without referring to the original data graph, path expressions of length of at most k, where k controls the resolution of the index and influences its size in a proportional manner. However, for large values of k, index size may still become very large. For small values of k, index size is substantially smaller, but the index cannot handle long path expressions.

---

[2] K-bisimilarity groups nodes with respect to local structure, i.e., the incoming paths of length up to k.

To accommodate path expressions of various lengths, without unnecessarily increasing the size of the whole index, D(k)-index assigns different values of k to different index nodes (Qun et al., 2003). These values conform to a given set of frequently-used path expressions (FUPs). Large values of k are assigned to parts of the index corresponding to parts of the data graph targeted by long path expressions; while small values of k are assigned to parts of the index corresponding to data targeted by short path expressions. To facilitate the evaluation of path expressions with branching, a variant called UD(k, l)-index also imposes downward similarity (Wu et al., 2003).

APEX is an adaptive index that searches for a trade-off between size and effectiveness (Chang et al., 2002). Instead of indexing all the paths from the root, APEX only indexes FUPs and preserves the source data structure in a tree. However, since FUPs are stored in the index, path query processing is quite efficient. APEX is also workload-aware, i.e., it can be dynamically updated according to changes in the query workload. A data mining method is used to extract FUPs from the workload for incremental update (Agrawal & Srikant, 1995).

Unfortunately, all these indexing techniques are ill-suited to decision-support queries. Data structures such as Dataguide, 1-index and its variants, and APEX are indeed applicable only on XML data that are targeted by simple path expressions. However, in the context of XML data warehouses, queries are complex and include several path expressions that compute join operations. Moreover, these indexes operate on one XML document only, whereas in XML warehouses, data are managed in several XML documents and decision-support queries are performed over these documents.

Finally, other techniques such as extended inverted lists (Zhang et al., 2001) and Fabric (Cooper et al., 2001) process containment queries over XML data stored in relational databases. Extended inverted lists includes a text index (T-index; Milo & Suciu, 1999) that is similar to traditional indexes in information retrieval systems, and an element index (E-index) that maps elements into inverted lists. Fabric indexes several XML documents by encoding paths, from root to leaf nodes, with indicators that code path labels. These codes are then inserted in a Patricia trie (Cooper et al., 2001) that processes them like simple characters. However, Fabric is not adapted to XML data warehouses either, because it does not take into account the relationships that exist between XML documents in a warehouse (facts and dimensions). This index is thus not beneficial to decision-support queries.

## Materialized view selection

The view selection problem has received significant attention in the literature, in the relational database context. To the best of our knowledge, no such view materialization approach exists in XML databases and XML data warehouses in particular. Researches about it differ by several points:

1. the way of determining candidate views;
2. the frameworks used to capture relationships between candidate views;
3. the use of mathematical cost models vs. calls to the query optimizer;
4. the context of view selection (relational vs. multidimensional);
5. the way optimization is performed (over multiple or single queries);

6. the nature of solutions (theoretical or technical).

Uchiyama et al. (1999) and Kotidis et al. (1999) introduce a lattice framework that models and captures dependency (ancestor or descendent) among aggregate views in a multidimensional context. This lattice is greedily browsed with the help of cost models to select the best views to materialize. This approach has also been used in one data cube and then extended to multiple cubes (Shukla et al., 2000). Valluri et al. (2002) propose another theoretical framework, AND-OR view graphs, to capture relationships between views. Though conceptually nice, these theoretical solutions are not truly scalable.

Most recent approaches are workload-driven. They syntactically analyze the query workload to enumerate relevant candidate views. A representative workload helps predict future queries, which are likely to belong to it or be syntactically close to current queries. Thus, extracting candidate materialized views from the workload ensures that they will probably be exploited when processing future queries. By calling to the DBMS' query optimizer (Agrawal et al., 2000) or by using maintenance cost and/or storage space constraints (Kotidis et al., 1999), workload-driven approaches greedily build a configuration of the most pertinent views. Clustering algorithms are also used to select pertinent views to materialize. Aouiche et al. (2006) indeed cluster similar queries together, and then merge queries in each cluster to build a set of candidate views. A greedy algorithm guided by cost models (for data access and storage) finally helps select the final set of views to materialize. In opposition to previous proposals, this approach is scalable thanks to the low complexity of clustering.

## STUDY CONTEXT

Although XML data warehouse architectures from the literature share a lot of concepts (mostly originating from classical data warehousing), they are nonetheless all different. Hence, we present in this section the XML data warehouse model that we choose and on which we base our query performance optimization techniques. We also present a sample XML decision-support query.

### XML data warehouse specification

When designing and building XML data warehouses, XML documents are used to manage or represent facts and/or dimensions. This allows natively storing documents and easily interrogating them with XML query languages.

Some XML warehousing approaches are user-driven. They are applied when an organization has fixed warehouse requirements. Nassis et al. (2005) propose methods to conceptually design and build an XML repository, based on object-oriented concepts and a view-driven approach, respectively. This repository represents the warehouse analysis context. Baril & Bellahsène (2003) envisage XML data warehouses as collections of views represented by XML documents. Zhang et al. (2005) propose an approach to materialize XML data warehouses based on the frequent query patterns discovered from historical queries issued by users. Finally, Vrdoljak et al. (2003) propose a design approach for Web warehouses that is based on XML schemas describing data sources. All these approaches assume that the warehouse is composed of XML documents representing facts.

Other approaches are explicitly based on classical data warehouse logical models. For instance, Pokorný (2003) models a star schema in XML by defining dimension hierarchies as sets of logically connected collections of XML data, and facts as XML data elements. Park et al. (2005) propose an XML multidimensional model in which each fact is described by a single XML document, and dimension data are grouped into a repository of XML documents. Rusu et al. (2005) build facts and dimensions from XML documents generated through XQueries. Eventually, Hümmer et al. (2003) propose a family of templates, called XCube, to describe a multidimensional structure (dimension and fact data) for integrating several data warehouses into a virtual or federated data warehouse. All these approaches assume that the warehouse is composed of XML documents that represent both facts and dimensions. They are used when dimensions are dynamic and allow the support of end-user analytical tools.

All these studies more or less converge toward a unified XML warehouse model. They mostly differ in the way dimensions are handled and the number of XML documents that are used to store facts and dimensions. In this chapter, we select the XCube specification, which is the most explicit, to model a reference XML data warehouse. However, since other models from the literature are quite similar, this is not a binding choice.

The advantage of XCube is its simple structure for representing facts and dimensions in a star schema. One XML document is used to represent dimensions and another one to represent facts. Hence, our reference data warehouse is composed of the following XML documents: *Schema.xml* specifies data warehouse metadata; *Dimensions.xml* defines all dimensions, each characterized by attributes their values; and *Facts.xml* specifies facts, i.e., sets of dimension identifiers and measure descriptions and values.

The tree structure of *Facts.xml* is described in Figure 1(a). Root node *CubeFact* has one child, *cube*, which is itself composed of *Cell* nodes defining facts, i.e., *fact* nodes (measures) and dimension references. A *fact* node has two attributes, *@id* and *@value, which* define the measure's name and value, respectively. A *dimension* node has two attributes, *@id* and *@value, which* define the dimension's name and its identifier's value, respectively.

The tree structure of *Dimensions.xml* is described in Figure 1(b). Root node *dimensionData* has one child, *classification*, which is itself composed of *Level* nodes. A *Level* node is composed of *node* nodes defining dimension instances. A *node* is composed of *attribute* nodes that define a dimension's attributes (*@name*) and their values (*@value*).

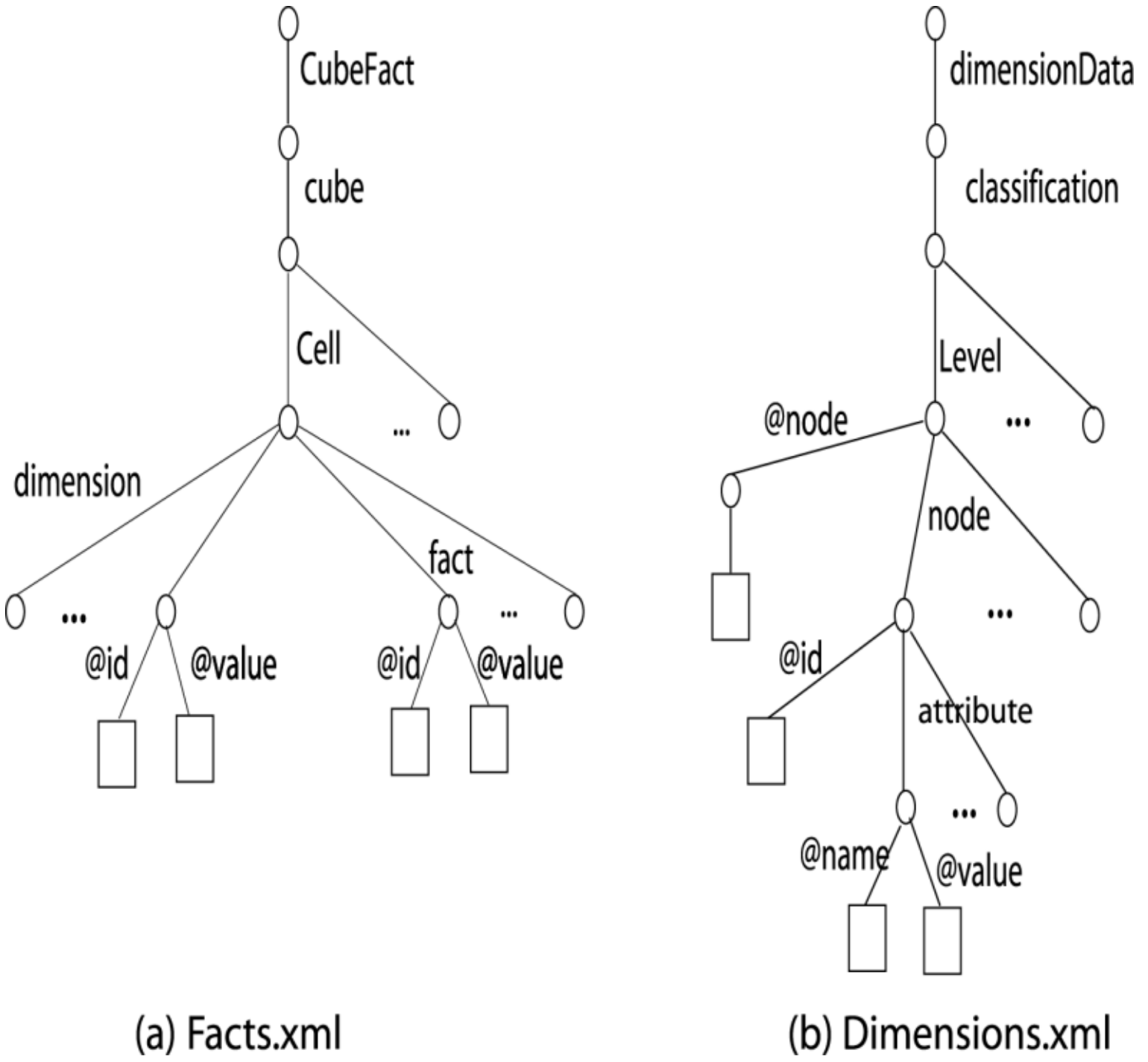

*Figure 1: Structure of dimension and fact documents*

## XML data warehouse interrogation

We select the XQuery language (Boag et al., 2004) to formulate decision-support queries because, unlike simpler languages such as XPath, it allows complex queries, including join queries over multiple XML documents, to be expressed with the FLWOR syntax.

However, XQuery does not support well the type of queries that are common in business analysis (Beyer et al., 2004). XQuery does indeed not include an explicit grouping construct comparable to the *group by* clause in SQL. Hence, several papers propose to extend XQuery to formulate decision-oriented queries (Borkar & Carey, 2004; Beyer et al., 2005). In our implementation, we acknowledge this effort by adding to FLWOR expressions explicit group by clauses. More precisely, we added two functions: *group by (attribute-list)* and *aggregation (aggregation-operations, measure-list)*, to the XQuery syntax. Figure 2 provides an example of decision-support query with a multiple *group by* clause that exploits these functions.

```
for $a in //dimensionData/classification/Level
[@node='customers']/node,
$x in //CubeFacts/cube/Cell
let $q := $b/attribute[@name='cust name']/@value
let $z := $b/attribute[@name='cust zip code']/@value
where $a/attribute/@name='cust city'
and $a/attribute/@value='Lyon'
and $x/dimension /@id=$a/@id
and $x/dimension/@id='customers'
group by(cust name,@cust zip code)
return name='cust name', aggregation(sum, quantity)
```

*Figure 2: Sample decision-support XQuery*

## JOIN INDEX FOR XML DATA WAREHOUSES

In this section, we present our join index structure and show how it allows indexing several related XML documents, which classical XML database indexes fail to do. We also present a theoretical study that to demonstrate our index' effectiveness.

### Join index structure

Building actual XML indexes on an XML warehouse causes a loss of information in decision-support query resolution. Indeed, clustering (1-index) or merging (Dataguide) identical labels causes the disappearance of relationships between fact measures and dimensions. We illustrate this problem in the following example.

The *Facts.xml* document is composed of *Cell* elements, each cell being characterized by dimension identifiers and one or more measures. Figure 3 shows the structures of *Facts.xml* and of its corresponding 1-index (which we selected as an example). 1-index represents cells linearly, i.e., all labels for the same source are represented by only one label. Hence, recovering a cell characterized by its measures and their dimension identifiers is impossible.

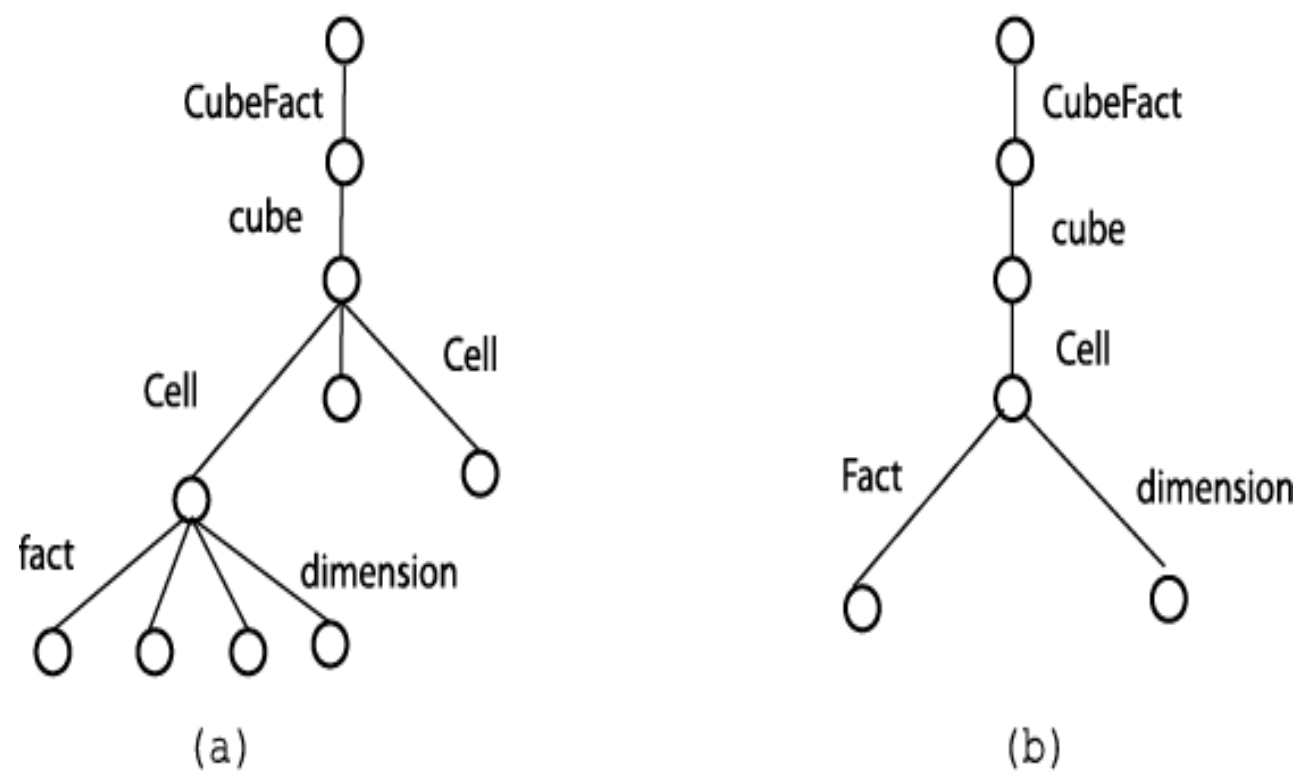

*Figure 3: Facts.xml structure (a) and corresponding 1-index (b)*

An index should be able to preserve the relationships between dimensions and fact measures. Thus, our index' structure is similar to that of the *Facts.xml* document, except for the attribute element. Moreover, XML indexes usually summarize or reorganize the structure of the indexed XML documents into new XML documents that are then accessed instead of original data. Our index structure is similar. It is stored in an XML document named *Index.xml*, whose structure is showed in Figure 4. Each *Cell* element is composed of dimensions and one or more facts. A *Fact* element has two attributes, *@id* and *@value*, which respectively represent measure names and values. Each *dimension* element is composed of two attributes: *@id*, which stores the dimension's name, and *@node*, which stores the dimension identifier's value. A *dimension* element also has children *attribute* elements. They are obtained from the *Dimensions.xml* document. An *attribute* element is composed of two attributes, *@name* and *@value*, which respectively store its name and value.

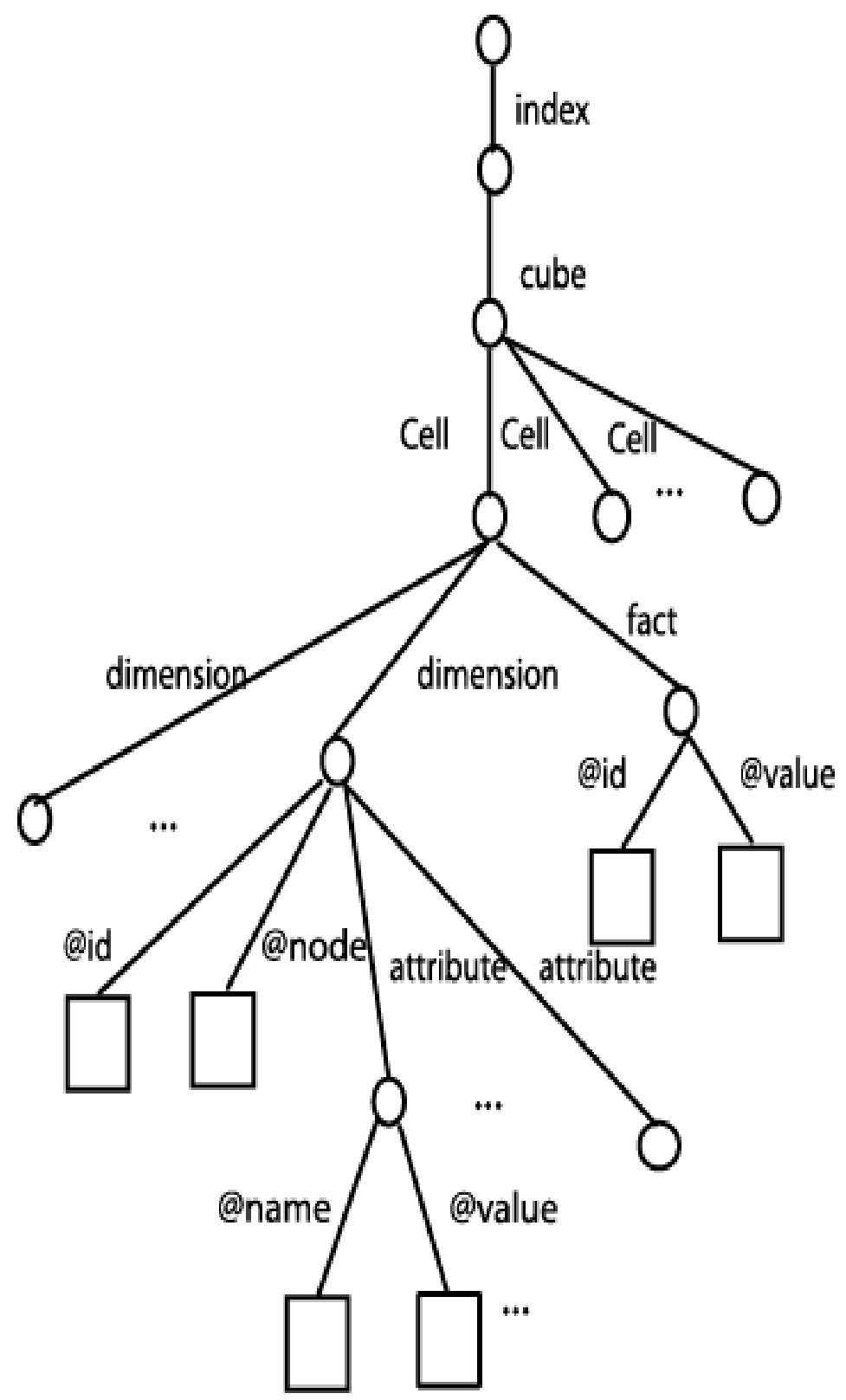

*Figure 4: XML join index structure*

Data migration from *Dimensions.xml* and *Facts.xml* to *Index.xml* helps store facts, dimensions and their attributes in the same cell. This feature wholly eliminates join operations since all the information that is necessary for a join operation is stored in the same cell. Queries need to be rewritten to exploit our index, though. The rewriting process consists in preserving selection expressions and aggregation operations. We illustrate query execution by an example in Figure 5. Further details are provided in the experiment section.

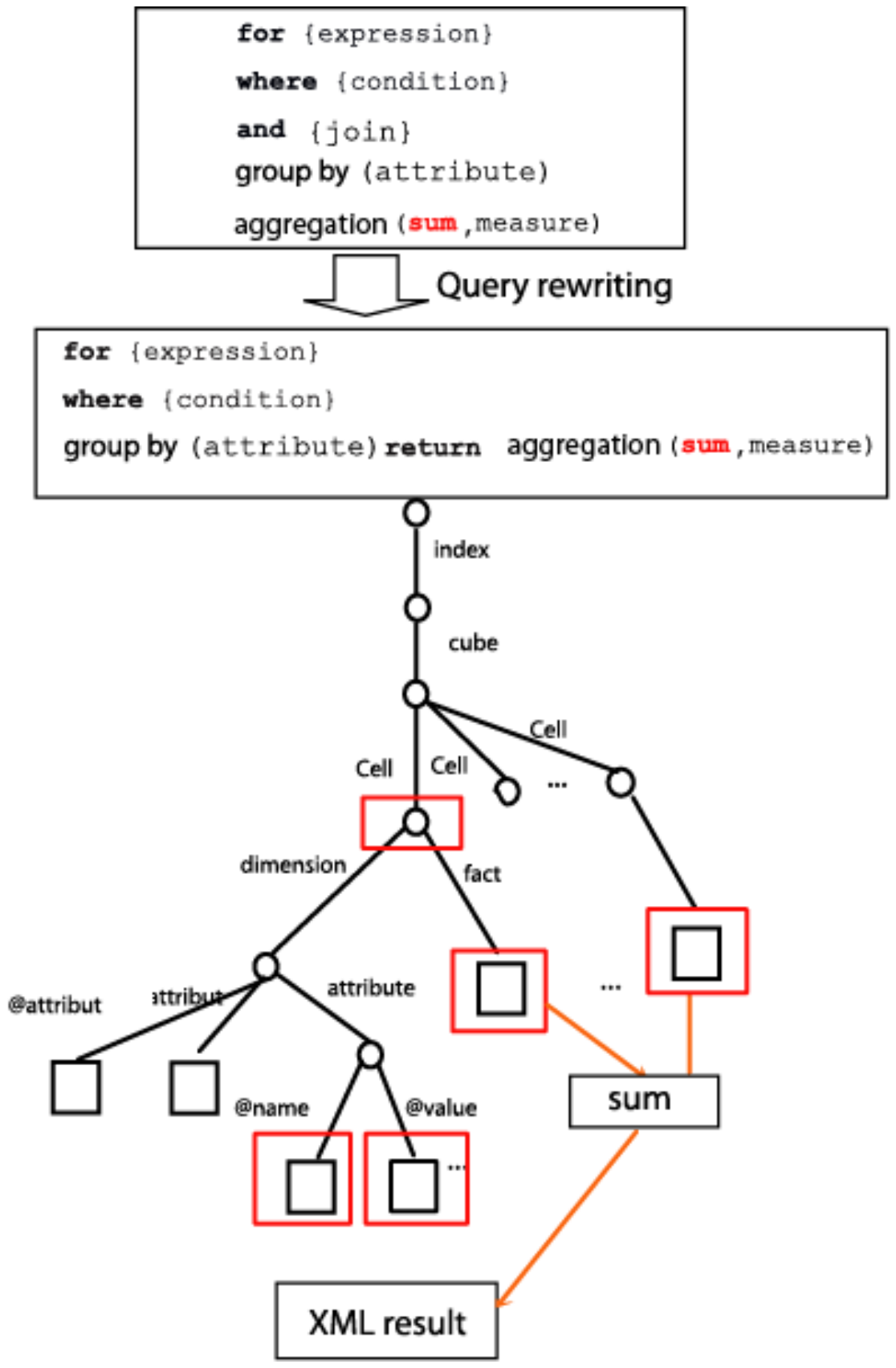

*Figure 5: Query executions over our join index*

## Theoretical validation

Queries defined over an XML warehouse modeled according to the XCube specification perform several join operations between facts stored in *Facts.xml* and dimensions from *Dimensions.xml*. Thus, they must satisfy the following constraints.

document(*Facts.xml*)/CubeFact/cube/Cell/dimension[@id =
document(*Dimensions.xml*)/classification/Level/@id]
and
document(*Facts.xml*)/CubeFact/cube/Cell/dimension/[@value =
document(*Dimension.xml*)/classification/Level/node/@id]

The first equality checks whether the dimension composing a cell (fact) is indeed the dimension expressed in the query. The second equality checks whether the node of a dimension (equivalent to a primary key) corresponds (can be joined) to the node, from the same dimension, defined in a cell (equivalent to a foreign key in the fact table).

Query execution without using our index may proceed as follows. For each dimension defined by *@node='name of dimension'*, identifiers *@id* verifying the *Where* clause are searched for. *Dimensions.xml* is traversed in depth first, down to the *Level* node. Child nodes of the *Level* node are then traversed in breadth first until *@node* is equal to dimension name. The cost of this traversal cost is equal to the number of *Level* nodes in *Dimensions.xml*, denoted |dimension|. If several dimensions are defined in the query, all *Level* nodes are traversed for each dimension. Each node's child is traversed in depth first, until a list of *@id* attributes verifying the conditions *@name='name of the attribute'* and *@value='value of the attribute'* is found. The cost of this traversal is equal to the number of *attribute* children. Thus, dimension cost traversal equals $|a_i| * |d_i|$, where $|a_i|$ is the number of attributes in each dimension and $|d_i|$ the number of *node* elements, i.e., the number of children in each dimension.

To join dimensions from *Dimensions.xml* and facts from *Facts.xml*, *@id* values found when processing dimensions are searched for in facts. *Facts.xml* is then traversed in depth first, down to the *Cell* level. Cells are then traversed in breadth first until dimensions whose child *@id* equals *@node* in *Dimensions.xml* and *@node* equals *@id* in *Dimensions.xml* are found. The traversal cost of *Facts.xml* is $|cell|$, where $|cell|$ is the number of cells. Finally, query execution cost without our index is defined by Formula 1.

$$E_{noindex} = \left( \left( |cell| * |dimension| \right) * \left( |dimension| + \left( |d_i| * |a_i| \right) \right) \right) \quad (1)$$

Query execution when using our index may proceed as follows. For each dimension defined by *@node='name of dimension'*, identifiers *@id* verifying the *Where* clause are searched for. *Index.xml* is traversed in depth first, down to the *Cell* level. The cost of this traversal is equal to the number of cells in *Index.xml*. Dimension child nodes are then traversed until the node whose *@id* value equals dimension name in the query is reached. The cost of this traversal is equal to the number of *dimension* nodes in *Index.xml*, i.e., the number of dimensions in the warehouse schema. This cost is denoted |*dimension*|. The children of each found node are traversed in depth first, down to the *attribute* node verifying the conditions *@name='name of the attribute'* and *@value='value of the attribute'*. The cost of this traversal is equal to the number of *attribute* children, denoted $|a_i|$. Finally, query processing cost over our index structure is defined by Formula 2.

$$E_{index} = |cell| * \left( |dimension| + |a_i| \right) \quad (2)$$

Figure 6 shows the cost variation between $E_{noindex}$ and $E_{index}$ with respect to the number of cells (facts) from *Facts.xml*. These facts are described by five dimensions that are stored in *Dimensions.xml*. Table 1 displays the characteristics of these dimensions. We use a logarithmic scale on the Y axis to better visualize cost differences. Using our index induces a performance gain factor of 14,000 on an average.

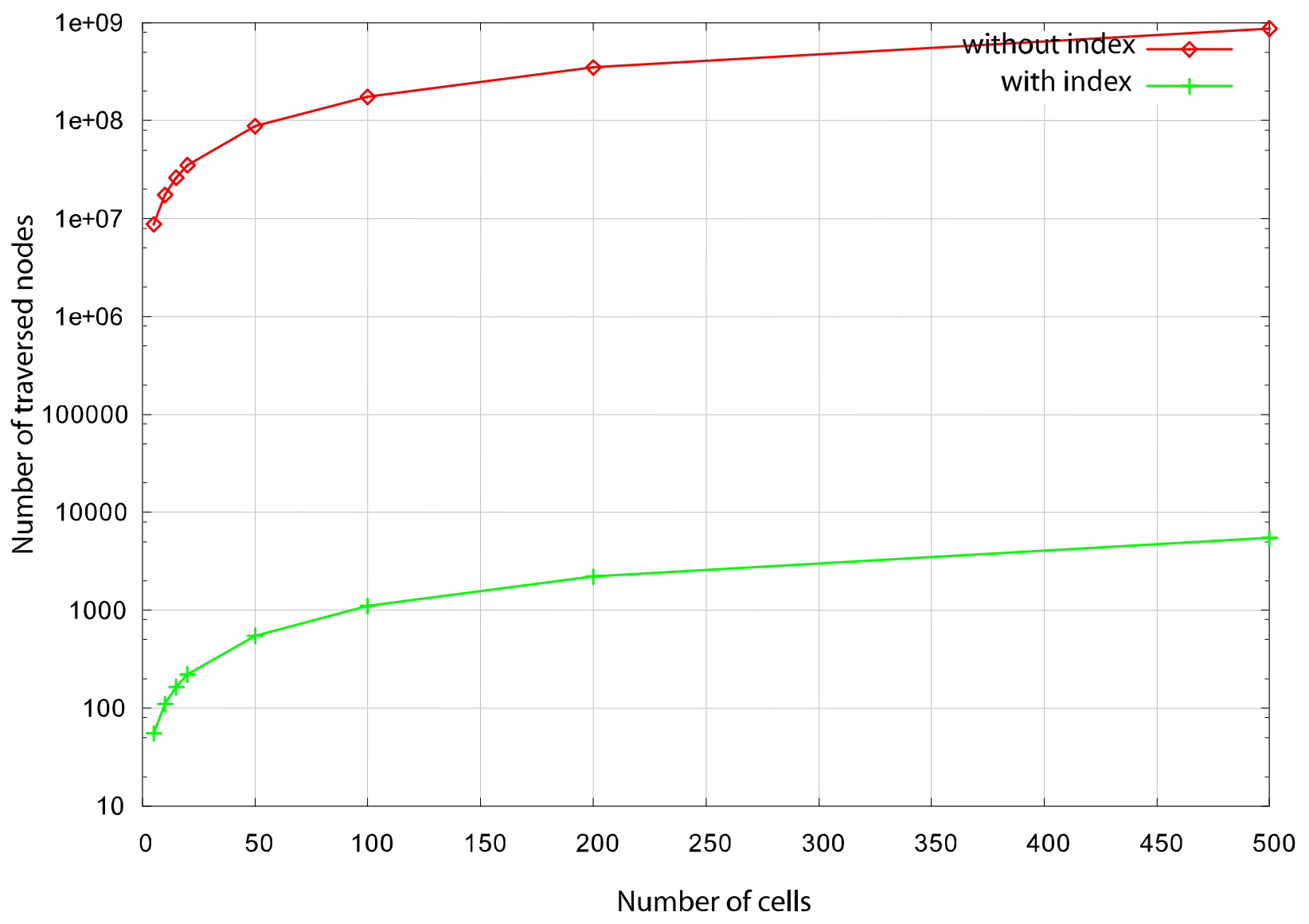

*Figure 6: Join index efficiency*

## XML MATERIALIZED VIEW SELECTION STRATEGY

In this section, we present our second contribution: an XML materialized view selection strategy. This strategy only materializes pertinent views, and hence addresses both storage and maintenance cost problems. It is workload-based (i.e., based on a set of queries representing user requirements) and exploits knowledge about how views can be used to resolve a set of workload queries to cluster them together.

The principle of our materialized view selection strategy is depicted in Figure 7. We assume that we dispose of a workload composed of representative queries (similar to the query from Figure 2). Our objective is to select a configuration of materialized views that reduces its execution time. The first step is to build, from the workload, a clustering context. Then, we define similarity and dissimilarity measures that help cluster together similar queries. For each cluster, we build a set of candidate views. The last step exploits cost models that evaluate the cost of accessing data using views and the cost of their storage, to build a final materialized view configuration. We detail these steps in the following sections.

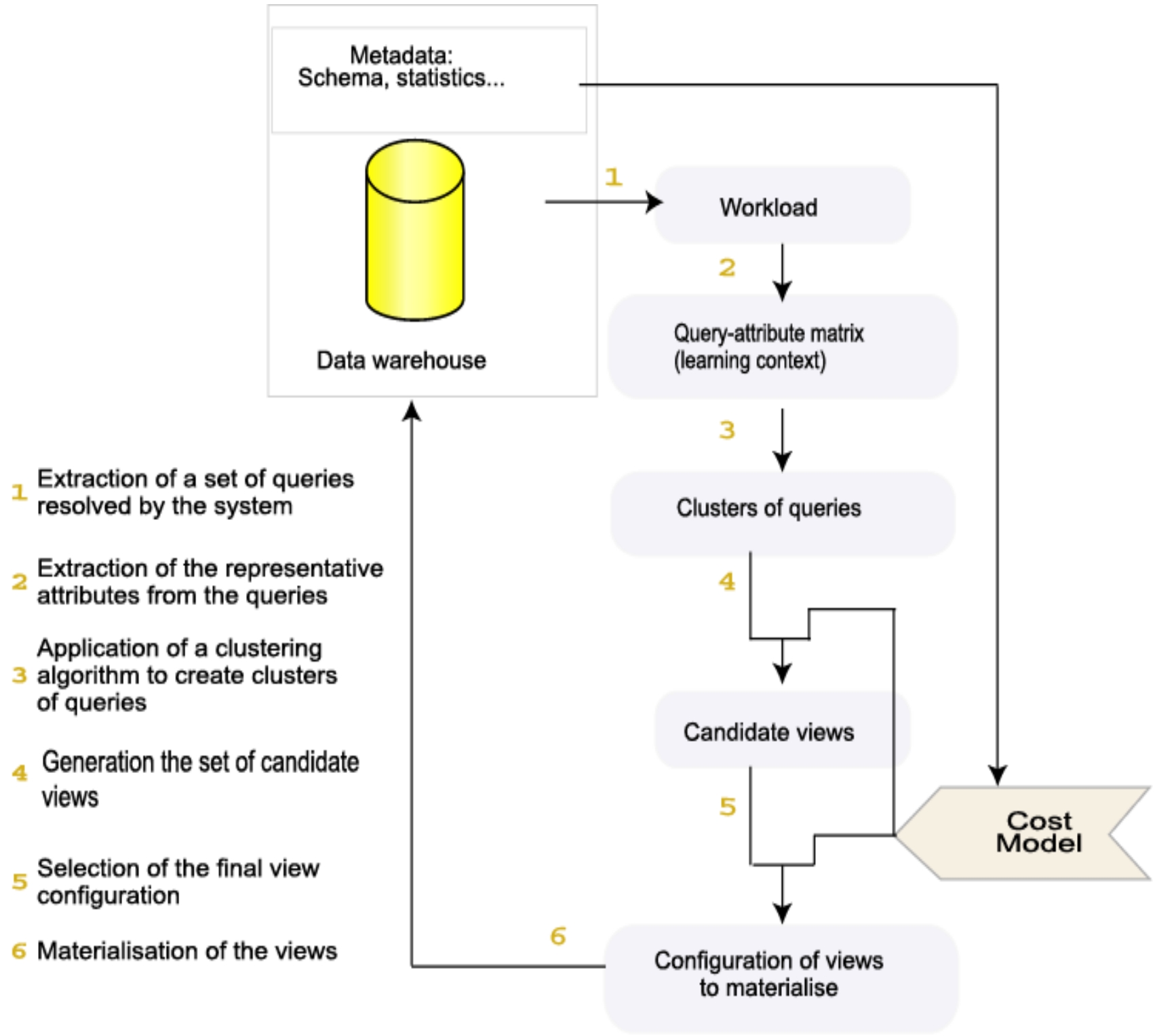

*Figure 7 Materialized view selection strategy*

## Query workload analysis

The workload we consider is a set of *selection*, *join* and *aggregation* queries. This first step consists in extracting from the workload representative attributes for each query. We mean by representative attributes those are present in *Where* (selection predicate attributes) and *Group by* clauses. We store the relationships between workload queries and the extracted attributes in a so-called "query-attribute" matrix. Matrix lines are queries and columns are extracted attributes. A query $q_i$ is then seen as a line in the matrix that is composed of cells corresponding to representative attributes. The general term $q_{ij}$ of this matrix is set to one if extracted attribute $a_i$ is present in query $q_i$, and to zero otherwise. This matrix represents our clustering context.

## Building the candidate view configuration

In practice, it is hard to search all the syntactically relevant candidate views because the search space is very large (Agrawal et al., 2000). To reduce the problem's size, we propose to cluster workload queries. Hence, we group in a same cluster all the queries that are similar. Similar queries are the one having a close binary representation in the query-attribute matrix. Two similar queries can be resolved by using only one materialized view. We define similarity and dissimilarity measures that ensure that queries within a same cluster are strongly related to each other, whereas queries from different clusters are significantly different from one another.

## Similarity and dissimilarity measures

A query is described by the attributes extracted in the query analysis phase. We thus describe a query $q_i$ by vector $q_i = \{q_{1i}, q_{1i}, ..., q_{pi}\}$, where *p* is the number of attributes in the matrix. This description allows query comparison. We define similarity (respectively, dissimilarity) between two queries $q_i$ and $q_j$ with respect to attribute $a_k (k = \overline{1..p})$ in Formula 1 (respectively, Formula 2).

$$\delta_{sim}(q_{ki}, q_{kj}) = \begin{cases} 1 \text{ if } q_{ki} = q_{kj} = 1 \\ 0 \text{ otherwise} \end{cases} \qquad (3)$$

$$\delta_{dissim}(q_{ki}, q_{kj}) = \begin{cases} 1 \text{ if } q_{ki} = q_{kj} \\ 0 \text{ if } q_{ki} \neq q_{kj} \end{cases} \qquad (4)$$

Two queries $q_i$ and $q_j$ are similar with respect to attribute $a_k$ if and only if $q_{ki} = q_{kj} = 1$, *i.e.*, $a_k$ is present in both queries. They are dissimilar if and only if $q_{ki} \neq q_{kj}$, *i.e.*, one of the two queries does not contain attribute $a_k$.

These measures can be extended to a set *A* composed of *p* attributes such that we get the degree of global similarity and dissimilarity between two queries. We thus define the similarity (respectively, dissimilarity) between two queries $q_i$ and $q_j$ with respect to all attributes in Formula 5 (respectively, Formula 6).

$$sim(q_i, q_j) = \sum_{j=1}^{p} \delta_{sim}(q_{ki}, q_{kj}) \qquad (5)$$
$$0 \leq sim(q_i, q_j) \leq p$$

$$dissim(q_i, q_j) = \sum_{j=1}^{p} \delta_{dissim}(q_{ki}, q_{kj}) \qquad (6)$$
$$0 \leq dissim(q_i, q_j) \leq p$$

Thus, the closer $sim(q_i, q_j)$ (respectively, $dissim(q_i, q_j)$) is to *p*, the more $q_i$ and $q_j$ are considered similar (respectively, dissimilar). We also define similarity (respectively, dissimilarity) measures between two query sets and within a query set. These measures are defined by Formulas 7, 8, 9 and 10, respectively.

$$sim(C_a, C_b) = \sum_{q_k \in C_a, q_l \in C_b} \delta_{sim}(q_k, q_l) \qquad (7)$$
$$0 \leq sim(C_a, C_b) \leq card(C_a) \times card(C_b) \times p$$

$$dissim(C_a, C_b) = \sum_{q_k \in C_a, q_l \in C_b} \delta_{dissim}(q_k, q_l) \qquad (8)$$

$$0 \leq dissim(C_a, C_b) \leq card(C_a) \times card(C_b) \times p$$

$$sim(C_a) = \sum_{q_k \in C_a, q_l \in C_b, k<1} \delta_{sim}(q_k, q_l) \qquad (9)$$

$$0 \leq sim(C_a, C_b) \leq \frac{card(C_a) \times card(C_b) \times p}{2}$$

$$dissim(C_a) = \sum_{q_k \in C_a, q_l \in C_b, k<1} \delta_{dissim}(q_k, q_l) \qquad (10)$$

$$0 \leq dissim(C_a, C_b) \leq \frac{card(C_a) \times card(C_b) \times p}{2}$$

## Clustering

Clustering consists in determining a so-called natural partition $P_{nat}$ composed of objects (here, queries) that reflect the internal structure of data. This partition must be such as its clusters are composed of objects with high similarity and objects from different clusters present a high dissimilarity. Based on the previously defined functions, a clustering quality measure $Q(P_h)$ can be built (Formula 11).

$$Q(P_h) = \sum_{a=1..z, b=1..z, a<b} (sim(C_a, C_b) + \sum_{a=1}^{z} dissim(C_a)) \qquad (11)$$

This measure permits to capture the natural aspect of a partition. Hence, $Q(P_h)$ measures simultaneously similarities between queries within the same cluster of partition $P_h$ and dissimilarities between queries within different clusters. Thus, we can define $Q(P_h)$ as a homogeneity function for the same class and a heterogeneity function for different classes. Therefore, partitions presenting high intra-cluster homogeneity and a high inter-cluster disparity have a low $Q(P_h)$ value and thereby appear as the most natural.

We have selected the Kerouac clustering algorithm (Jouve & Nicoloyannis, 2003). Kerouac indeed bears several interesting properties:

1. its computational complexity is low (log linear with respect to the number of queries and linear with respect to the number of attributes);
2. it can deal with a high number of objects (queries);
3. it can deal with distributed data;
4. it allows integrating constraints within the clustering process. This last characteristic is particularly interesting, since it provides us with a way to integrate constraints concerning the view merging process.

## Cost models

The number of candidate views is generally as high as the input workload is large. Thus, it is not feasible to materialize all the proposed views because of storage space constraints. To circumvent this limitation, we propose to use cost models that retain the most pertinent views only.

Figure 8 shows the typical structure of an XML view. In our context, it is composed of *Cell* elements. Each *Cell* is itself composed of *Dimension* elements that contain *Group by* attributes and *Fact* elements corresponding to aggregate results. We propose cost models that estimate the size and storage cost of a given XML view.

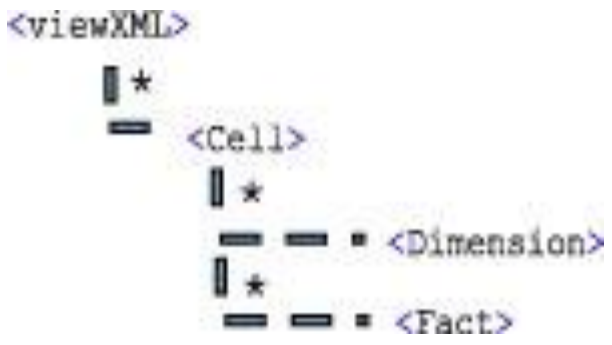

*Figure 8: XML view structure*

We estimate the size of a view by its number of elements. The number of *Dimension* and *Fact* elements in each *Cell* is the same. Indeed, the number of elements in a given view is estimated by the number of *Cell* elements. To compute it, we first estimate the maximum number of *Cell* elements (Formula 12).

$$ms(Cell) = \prod_{i=1}^{d} |d_i| \qquad (12)$$

$|d_i|$ is the cardinality of the dimension characterizing the *Cell* element. *d* is the number of dimensions in *Dimensions.xml*. Let *ms*(*v*) be the maximum size of view *v* that is composed of dimensions $d_1 \ldots d_k$, where *k* is the number of dimensions in the view and $|d_i|$ the cardinality of dimension $d_i$ (Formula 13).

$$ms(v) = \prod_{i=1}^{k} |d_i| \qquad (13)$$

Golfarelli & Rizzi (1998) proposed to estimate the number of tuples in a given view *v* by using Yao's (1977) formula. We also use this formula to estimate the number of *Cell* elements in *v* (Formula 14).

$$|v| = ms(v) \times \left[ 1 - \prod_{i=1}^{Cell} \frac{ms(Cell) \times c - i + 1}{ms(Cell) - i + 1} \right] \qquad (14)$$

$c = 1 - \frac{1}{ms(v)}$. If $\frac{ms(Cell)}{ms(v)}$ is large enough, this formula is well approximated by Cardenas' (1975) formula (Formula 15).

$$|v| = ms(v) \times \left(1 - \left(1 - \frac{1}{ms(v)}\right)^{Cell}\right) \qquad (15)$$

Cardenas and Yao's formulas are based on the assumption that data are uniformly distributed. The size, in bytes, of a view *v* is equal to the number of *Cell* elements multiplied by the average size needed to store one element. Thus, we estimate the size of a view as shown in Formula 16.

$$size(v) = |v| \times \sum_{i=1}^{k} size(d_i) \qquad (16)$$

$size(d_i)$ represents the size, in bytes, of dimension $d_i$ from *v* and *k* the number of dimensions.

## Objective functions

We describe in this section three objective functions that help evaluate the variation of query execution cost induced by adding a new view. Query execution cost is assimilated either to the number of *Cell* elements in *Facts.xml*, if no views are used; or to the number of *Cell* elements in the views if they are exploited. Workload execution cost is obtained by adding the execution costs of each query within this workload.

The first objective function, "profit", favors views providing more profit while executing queries. The second function, "profit/space ratio", favors views providing more benefit while occupying the smallest possible storage space. The third function, "hybrid", combines the first two in order to first select all the views providing more profit and then retain only those occupying less storage space when this resource becomes critical. The profit function is useful when storage space is not limited, the profit/space ratio function is useful when storage space is small, and the hybrid function is interesting when storage space is reasonably large.

### Profit objective function

Let $V = \{v_1, \ldots, v_m\}$ be the candidate view set, *S* the final view set and $Q = \{q_1, \ldots, q_n\}$ a query set (workload). The profit objective function, noted *P*, is defined in Formula 17.

$$P_{/S}(v_j) = C_{/S}(Q) - C_{/S \cup vj}(Q) - \beta C_{update}(v_j) \qquad (17)$$
$$(v_j) \notin S$$

$C_{/S}(Q)$ is the query execution cost when all views in *S* are used. If this set is empty, $C_{/\text{Ø}}(Q) = |Q| \times |F|$ because all queries are resolved by accessing fact *F*. When a view $v_i$ is added to *S*, $C_{/S \cup vj}(Q) = \sum_{k=0}^{|Q|} C(q_k, v_j)$ is the query execution cost for views in $S \cup v_i$. If query $q_k$ exploits $v_i$, cost $C(q_k, v_j)$ is then equal to $C_{vj}$ (number of instances in $v_i$). Otherwise, $C(q_k, v_j)$ is equal to the maximum value between *F* and $C(q_k, v)$ (executing cost of

$q_i$ exploiting $v \in S$ with $v \neq v_j$). Coefficient $\beta = |Q| p(v_i)$ estimates the number of updates for view $v_i$. The update probability $p(v_i)$ equals $\frac{1}{storage-space} \frac{\%update}{\%query}$, where ratio $\frac{\%update}{\%query}$ represents the proportion of update vs. interrogation queries. Finally, $C_{update}(v_j)$ represents the maintenance cost of view $v_j$.

**Profit/space ratio objective function**

If view selection is achieved under a space constraint, the profit/space objective function from Formula 18 is used. This function *R* computes the profit provided by $v_j$ with respect to storage space $size(v_j)$ it occupies.

$$R_{/S}(v_j) = \frac{P_{/S}(v_j)}{size(v_j)} \qquad (18)$$

**Hybrid objective function**

The constraint on storage space may be relaxed if storage space in relatively large. The hybrid objective function *H* does not penalize space-greedy views if ratio $\frac{remaining-space}{storage-space}$ is lower or equal than a storage-space given threshold α, 0 < α ≤ 1, where *remaining-space* and *storage-space* are respectively the remaining space after adding $v_j$ and the allotted space needed for storing all the views. This function is computed by combining functions *P* and *R* as shown in Formula 19.

$$H_{/S}(v_j) = \begin{cases} P_{/S}(v_j) \text{ if } \frac{remaining-space}{storage-space} \leq \alpha \\ R_{/S}(v_j) \text{ otherwise} \end{cases} \qquad (19)$$

## View selection algorithm

Our view selection algorithm (Algorithm 1) is based on a greedy search within the candidate view set *V*. Objective function *F* must be one of among functions *P*, *R* or *H* described in the previous section. If *R* is used, we add to the algorithm's input the storage space *M* allotted for views. If *H* is used, we also add threshold *α* as input.

$$
\begin{array}{l}
S \leftarrow \varnothing \\
\textbf{repeat} \\
\quad v_{\max} \leftarrow \varnothing \\
\quad F_{\max} \leftarrow 0 \\
\quad \textbf{forall } v_j \in V - D \textbf{ do} \\
\qquad \textbf{if } F_{/S}(v_j) > F_{\max} \textbf{ then} \\
\qquad\quad F_{\max} \leftarrow F_{/S}(v_j) \\
\qquad\quad v_{\max} \leftarrow v_j \\
\qquad \textbf{endif} \\
\quad \textbf{endfor} \\
\quad \textbf{if } F_{/S}(v_j) > 0 \textbf{ then} \\
\qquad S \leftarrow S \cup \{v_{\max}\} \\
\quad \textbf{endif} \\
\textbf{until } (F_{/S}(v_j) \leq 0 \textbf{ or } V - S = \varnothing)
\end{array}
$$

*Algorithm 1: Greedy view configuration construction*

In the first iteration, the values of the objective function are computed for each view within *V*. The view $v_{\max}$ that minimizes *F*, if it exists ($F_{/S}(v_{\max}) > 0$), is then added to *S*. If *R* or *H* is used, the whole storage space *M* is decreased by the amount of space occupied by $v_{\max}$. Values of *F* are then computed for each remaining view in *V* − *S*, since they depend on the selected views present in *S*. This helps take into account the interactions that probably exist between views. This process reiterates either until there is no performance improvement ($F_{/S}(v) > 0$) or until all the views have been selected ($V - S = \varnothing$). If functions *R* or *H* are used, the algorithm also stops when storage space is full.

## VALIDATION EXPERIMENTS

### Experimental conditions

In order to validate our proposals experimentally, we exploit an XML data warehouse modeled according to the XCube specifications. Actual data have been transferred from an existing, relational data warehouse derived from an Oracle example (Oracle, 2006).This classical test data warehouse is modeled as a star schema composed of *sale* facts characterized by the *amount* (of purchased products) and *quantity* (of purchased products) measures. Facts are stored in the *Facts.xml* document. They are described by five dimensions: *channels*, *promotions*, *customers*, *products* and *times* that are stored in the *Dimensions.xml* document. Table 1 displays the characteristics of our test XML data warehouse.

We implemented this data warehouse within two native XML DBMSs: eXist (Meier, 2002) and X-Hive (Waldt, 2005). Both these DBMSs allow the native storage of large documents and support the XQuery language. They also provide APIs (Application Programming Interfaces) for storing, querying, retrieving, transforming and publishing XML data. We also implemented our XML data warehouse in a relational, XML-compatible DBMS: SQL Server (Rys, 2004). SQL Server 2005 handles XML data through an XML type field. It integrates XQuery queries with the help of a function called query that is embedded into SQL Select clauses (Figure 9).

| **Facts** | **Number of cells** |
|---|---|
| Sales | 16 260 336 |
| **Dimensions** | **Number of occurrences** |
| Customers | 50 000 |
| Products | 10 000 |
| Times | 1 461 |
| Promotions | 501 |
| Channels | 5 |
| **Documents** | **Size (MB)** |
| Facts.xml | 4.92 |
| Dimensions.xml | 3.77 |
| Schema.xml | 0.001 |

*Table 1: Test data warehouse characteristics*

```
Select XML-DOC.Query('for $a in //dimensionData
        /classification/Level[@node='customers']/node,
        where $a/attribute/@name="cust city"
        and $a/attribute/@value="Lyon"
        return name="cust name"')
From DIMENSION
```

*Figure 9: Sample SQL-XQuery*

## Join index evaluation

This experiment measure the execution time of the typical decision-support query from Figure 2 over our test data warehouse, with and without exploiting our join index, on all the DBMSs we consider (Figure 10). We also vary warehouse size. Note that, in SQL Server 2005, XML data are stored in a table field. Thus, SQL-XQuery queries must be processed for each record. This process does not allow joining XML data from different records. Hence, we only perform our experiment with our join index on SQL Server since it is not possible with the original, multi-document warehouse. We ran our tests on a Pentium 2 GHz PC with 1 GB of main memory and an IDE hard drive. Also note that we do not consider index construction time here, since an XML index is actually a new warehouse structure that is built once and queried thereafter. Finally, in Figure 10, the X axis represents warehouse size and the Y axis the corresponding execution time. The Y axis is in logarithm scale to highlight the differences in execution costs.

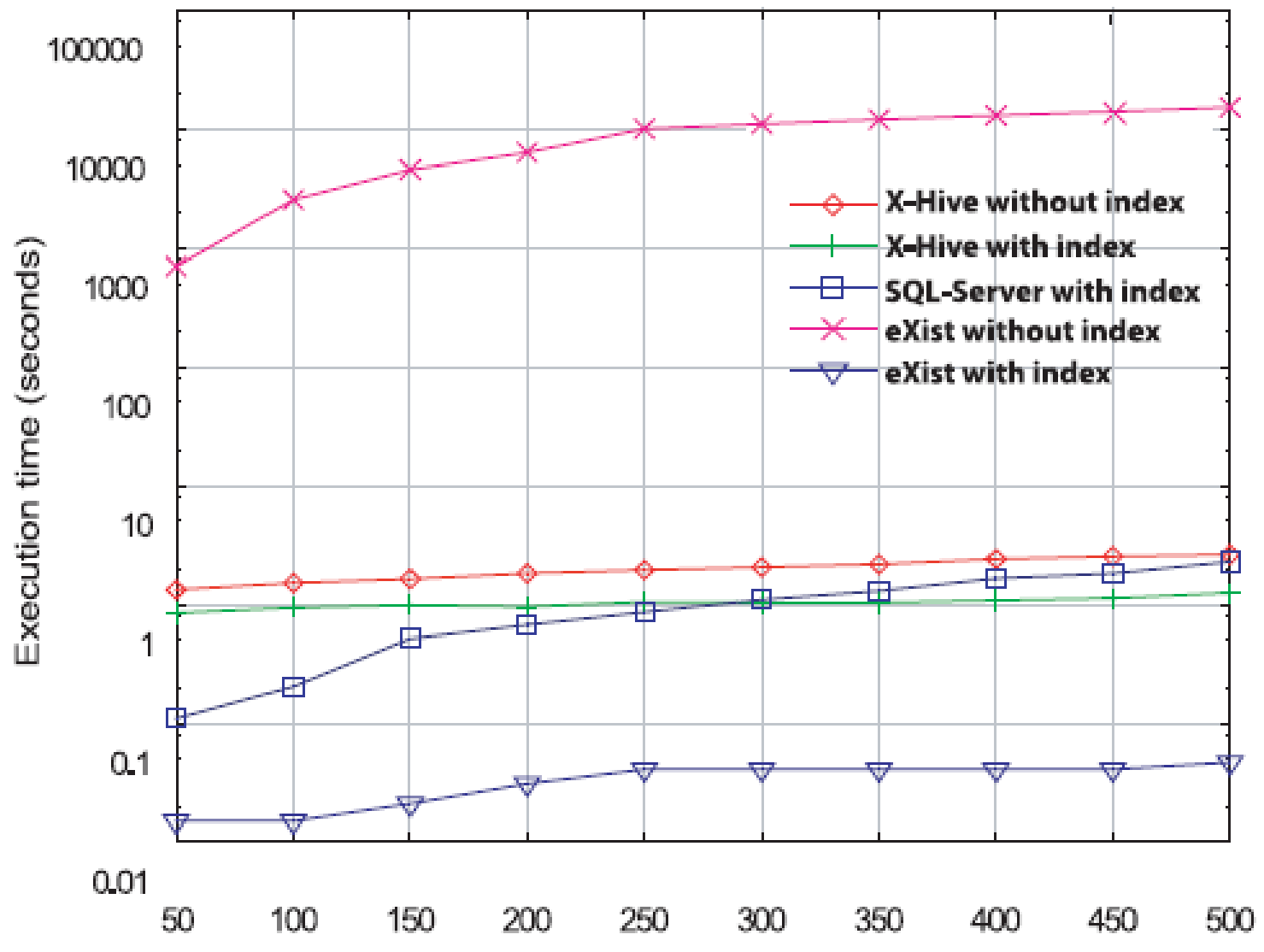

*Figure 10: Join index experimental results*

The results we obtain show that using our index structure significantly improves response time. On an average, the gain factor is indeed 25,669for eXist and 8,411 for X-Hive. Though this is not plotted on Figure 10, we also pushed our ”with/without index” tests further on the totality of the cells from *Facts.xml*. We achieve execution times of less than two seconds with our join index. Without index, X-Hive responds in about four minutes and eXist proves unable to answer in a reasonable time. Finally, this experiment shows that, properly indexed, native XML DBMSs can compete with, and even best relational DBMSs in terms of performance when XML documents are bulky. eXist running on our join index indeed outperforms SQL-Server by a 31.5 factor, on an average. This is because relational DBMS engines combine XQuery to SQL and must convert the result from relations to XML. XML native DBMSs, on the other hand, preserve the hierarchical structure of XML data, which allows path scans to be efficiently processed by XQuery engines. Our experiment also shows that eXist's query engine performs better then X-hive when using simple path expressions. We think this is because, eXist implements a specific numbering scheme that helps easily evaluate parent/child node relationships (Meier, 2002).

## Materialized view selection evaluation

To validate our materialized view selection strategy, we executed on our test data warehouse a workload composed of ten decision-support XQueries, with and without building materialized views. The selected views are stored in an independent collection that is targeted by rewritten queries. We plot in Figure 11 the execution time of our query workload on the original XML documents and on the materialized views we generate. The X-axis represents the ten queries and the Y-axis the corresponding execution time. The Y-axis is represented in logarithmic scale to highlight the difference between the execution costs. On an average, our XML view materializing strategy improves response time by a factor 24,700.

These results show that query response time significantly decreases using our strategy. Indeed, queries exploiting views obtained with our strategy are rewritten and join operations are avoided.

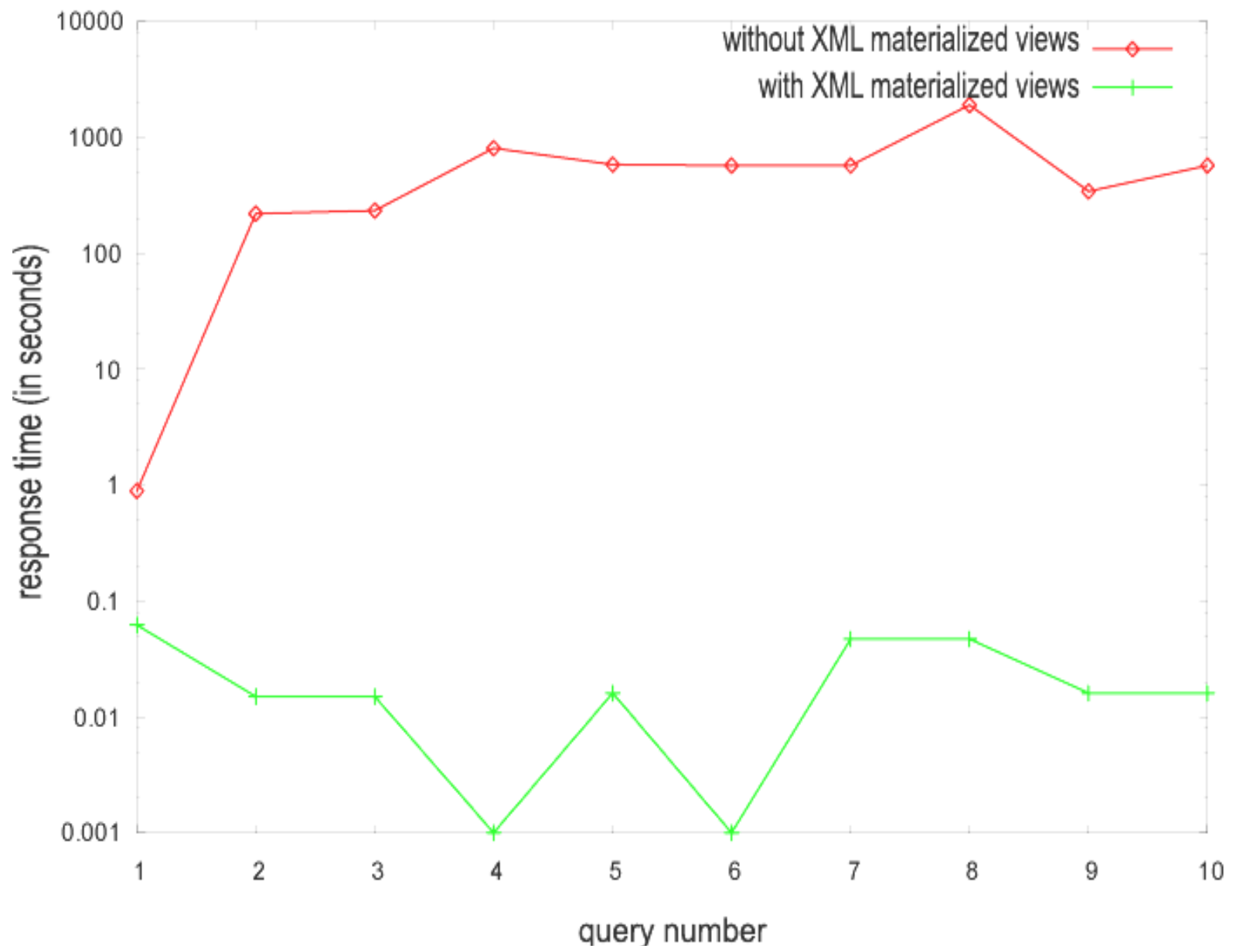

*Figure 11 Materialized view selection experimental results*

# CONCLUSION AND PERSPECTIVES

In this paper, we presented both a new join index and a strategy for materializing views in XML data warehouses. Our join index allows optimizing access time to several XML documents by eliminating join costs, while preserving the information contained in the initial warehouse. Our materialized view selection strategy exploits the results of clustering applied on a given workload to build a set of syntactically relevant candidate views. With the help of cost models we specifically designed for XML warehouses, we retain only the most advantageous candidate views through a greedy process that operates under storage space constraint.
To validate our join index, we performed both a complexity study and experiments. We implemented our reference warehouse with two native XML DBMSs and one relational, XML-compatible DBMS. Our tests showed that using our index structure significantly improves the response time of a typical decision-support query expressed in XQuery. Furthermore, they also demonstrate that native XML DBMSs can compete with relational DBMSs. The experimental results we achieved to validate our materialized view selection strategy are very encouraging, and show that it guarantees a substantial gain in performance. However, our first perspective is to complement these results with other tests, on other systems than eXist, and to assert in each configuration the gain in performance vs. the overhead for generating and refreshing materialized views.

This work also opens broader axes of research perspectives. First, our indexing strategy could be better integrated into a host native XML DBMS. This would certainly help develop an incremental strategy for the maintenance of the join index data structure. Moreover, the mechanism for rewriting queries would also be more efficient if it was part of the system. XML indexes are getting more and more efficient, but there is still room for improvement (e.g., multi-document join indexes). The generalized exploitation of materialized views would also be very beneficial. Thus, a rewriting query engine and refreshing strategies should be devised.

## REFERENCES

Agrawal, R. & Srikant, R. (1995). Mining Sequential Patterns. In: 11th International Conference on Data Engineering (ICDE 95), Taipei, Taiwan, IEEE Computer Society. pp. 3–14

Agrawal, S., Chaudhuri, S. & Narasayya, V. R. (2000). Automated selection of materialized views and indexes in SQL databases. In: 26th International Conference on Very Large Data Bases (VLDB 00), Cairo, Egypt. pp. 496–505.

Aouiche, K., Jouve, P.E. & Darmont, J. (2006). Clustering-Based Materialized View Selection in Data Warehouses. In : 10th East European Conference on Advances in Databases and Information Systems (ADBIS 06). Vol. 4152 LNCS Springer. pp. 81-95.

Baril, X. & Bellahsène, Z. (2003). XML Data Management: Native XML and XML-enabled Database Systems. In: Designing and Managing an XML Warehouse. Addison Wesley. pp. 455–473

Beyer, K. S., Chamberlin, D. D., Colby, L. S., Ozcan, F., Pirahesh, H. & Xu, Y. (2005). Extending XQuery for Analytics. In: ACM SIGMOD International Conference on Management of Data (SIGMOD’05), Baltimore, USA. ACM, 503–514.

Beyer, K.S., Cochrane, R., Colby, L.S., Ozcan, F. & Pirahesh, H. (2004). XQuery for Analytics: Challenges and Requirements. In: 1st International Workshop on XQuery Implementation, Experience and Perspectives <XIME-P/>, Paris, France. pp. 3–8

Boag, S., Chamberlin, D., Fernandez, M., Florescu, D., Robie, J. & Siméon, J. (2004). XQuery 1.0: An XML Query Language. W3C Working Draft, http://www.w3.org/TR/xquery/

Borkar, V. & Carey, M. (2004). Extending XQuery for Grouping, Duplicate Elimination, and Outer Join. In: <XML 2004> Conference & Exhibition, Washington DC, USA. pp. 1–11

Cardenas, A. F. (1975). Analysis and performance of inverted data base structures. Communications of the ACM, Vol. 18, No. 5, , pp. 253–263.

Chung, C. W., Park, M.J. & Shim, K. (2002). APEX: an Adaptive Path Index of XML data. In: ACM SIGMOD International Conference on Management of Data (SIGMOD 02). pp. 121-132.

Chung, C.W., Min, J.K. & Shim, K. (2002). APEX: an adaptive path index for XML data. In: ACM SIGMOD International Conference on Management of Data (SIGMOD 02), Madison, Wisconsin, ACM. pp. 121–132

Cooper, B.F., Sample, N., Franklin, M.J., Hjaltason, G.R. & Shadmon, M. (2001). A Fast Index for Semistructured Data. In: 27th International Conference on Very Large Data Bases (VLDB 01), Roma, Italy, Morgan Kaufmann. pp. 341–350

Darmont, J., Boussaïd, O., Bentayeb, F., Rabaseda, S., & Zellouf, Y. (2003). Web multiform data structuring for warehousing. In: Multimedia Systems and Applications. Vol. 22 Kluwer pp. 179–194

Darmont, J., Boussaïd, O., Ralaivao, J.C., & Aouiche, K. (2005). An Architecture Framework for Complex Data Warehouses. In: 7th International Conference on Enterprise Information Systems (ICEIS 05), Miami, USA. pp. 370–373

Goldman, R. & Widom, J. (1997) DataGuides: Enabling Query Formulation and Optimization in Semistructured Databases. In: 23rd International Conference on Very Large Data Bases (VLDB 97), Athens, Greece, Morgan Kaufmann pp. 436–445

Golfarelli, M. & Rizzi, S. (1998). A methodological framework for data warehouse design. In 1st ACM international workshop on Data warehousing and OLAP (DOLAP 1998), New York, USA, pp 3–9.

Gupta, H. & Mumick, I. S. (2005). Selection of Views to Materialize in a Data Warehouse. IEEE Transactions on Knowledge and Data Engineering, Vol. 17, No. 1. pp24–43.

Hümmer, W., Bauer, A. & Harde, G. (2003). XCube: XML for data warehouses. In: 6th International Workshop on Data Warehousing and OLAP (DOLAP 03), New Orleans, USA, ACM. pp. 33–40

Jouve, P. & Nicoloyannis, N. (2003). Kerouac: An algorithm for clustering categorical data sets with practical advantages. In International Workshop on Data Mining Learning for Actionable Knowledge (DMAK 2003).

Kaushik, R., Shenoy, P., Bohannon, P. & Gudes, E. (2002). Exploiting Local Similarity for Indexing Paths in Graph-Structured Data. In: 18th International Conference on Data Engineering (ICDE 02), San Jose, CA, IEEE Computer Society. pp. 129–140

Kimball, R., & Ross, M. (2002). The Data Warehouse Toolkit: The Complete Guide To Dimensional Modeling, 2nd Edition. John Wiley.

Kotidis, Y. & Roussopoulos, N. (1999). Dynamat: A dynamic view management system for data warehouses. In ACM SIGMOD International Conference on Management of Data (SIGMOD 1999), Philadelphia, USA, pp 371– 382.

Meier, W., (2002). eXist: An Open Source Native XML Database. In: Web, Web-Services, and Database Systems, NODe 2002 Web and Database-Related Workshops, Erfurt, Germany. Vol. 2593 of Lecture Notes in Computer Science. Springer, 169–183.

Milo, T. & Suciu, D. (1999). Index Structures for Path Expressions. In: 7th International Conference on Database Theory (ICDT 99), Jerusalem, Israel. Volume 1540 of Lecture Notes in Computer Science., Springer. pp. 277–295

Nassis, V., Rajugan, R., Dillon, T.S. & Rahayu, J.W. (2005). Conceptual and Systematic Design Approach for XML Document Warehouses. International Journal of Data Warehousing & Mining 1(3). pp. 63–86

Oracle Corporation. (2006). Oracle9i Data Warehousing Guide Release 2 (9.2). http://downloadwest.oracle.com/docs/cd/B10501 01/server.920/a96520/toc.htm

Park, B.K., Han, H. & Song, I.Y. (2005). XML-OLAP: A Multidimensional Analysis Framework for XML Warehouses. In: 7th International Conference on Data Warehousing and Knowledge Discovery, (DaWaK 05), Copenhagen, Denmark. Volume 3589 of Lecture Notes in Computer Science., Springer. pp. 32–42

Pokorný, J. (2002). XML Data Warehouse: Modelling and Querying. In: 5th International Baltic Conference (BalticDB&IS 02), Tallin, Estonia, Tallin Technical University. pp. 267–280

Qun, C., Lim, A. & Ong, K.W. (2003). D(k)-Index: An Adaptive Structural Summary for Graph-Structured Data. In: 2003 ACM SIGMOD International Conference on Management of Data (SIGMOD 03), San Diego, USA, ACM. pp. 134–144

Rusu, L.I., Rahayu, J.W. & Taniar, D. (2005). A Methodology for Building XML Data Warehouse. International Journal of Data Warehousing & Mining 1(2). pp. 67–92

Rys, M. (2004). XQuery in Relational Database Systems. In XML 2004 Conference and Exposition Proceedings (XML 2004), Washington, USA.

Shukla, A. Deshpande, P. & Naughton, J. F. (2000). Materialized view selection for multi-cube data models. In 7th International Conference on Extending DataBase Technology (EDBT 2000), Konstanz, Germany, pages 269-284.

Uchiyama, H., Runapongsa, K., & Teorey, T. J. (1999). A progressive view materialization algorithm. In 2nd ACM International Workshop on Data warehousing and OLAP (DOLAP 1999), Kansas City, USA, pp. 36–41.

Valluri, S. R., Vadapalli, S. & Karlapalem, K. (2002). View relevance driven materialized view selection in data warehousing environment. In 30th Australasian conference on Database technologies, Melbourne, Australia, pages 187-196.

Vrdoljak, B., Banek, M. & Rizzi, S. (2003). Designing Web Warehouses from XML Schemas. In: 5th International Conference on Data Warehousing and Knowledge Discovery (DaWaK 03), Prague, Czech Republic. Volume 2737 of Lecture Notes in Computer Science., Springer. pp. 89–98

Waldt, D. (2005). Using XML and Databases: W3C Standards in Practice. White Paper, The Gilbane Report, http://www.x-hive.com Wu, H., Wang, Q., Yu, J.X., Zhou, A. & Zhou, S. (2003). UD(k, l)-Index: An Efficient Approximate Index for XML Data. In: 4th International Conference on Advances in Web-Age Information Management (WAIM 03), Chengdu, China. Volume 2762 of Lecture Notes in Computer Science, Springer. pp. 68–79

Yao, S. (1997). Approximating block accesses in database organizations. Communication of the ACM, Vol. 20, No. 4, pp 260–261.

Zhang, C., Naughton, J.F., DeWitt, D.J., Luo, Q. & Lohman, G.M. (2001). On Supporting Containment Queries in Relational Database Management Systems. In: ACM SIGMOD International Conference on Management of Data (SIGMOD 01), Santa Barbara, USA, ACM. pp. 425–436

Zhang, J., Wang, W., Liu, H. & Zhang, S. (2005). X-warehouse: building query pattern driven data. In: 14th International Conference on World Wide Web (WWW 05), Chiba, Japan, ACM. pp. 896–897

## KEY TERMS & DEFINITIONS

Database management system (DBMS): software set that handles structuring, storage, maintenance, update and querying of data stored in a database.

XML-native DBMS (NXD): database system in which XML data are natively stored and queried as XML documents. An NXD provides XML schema storage and implements an XML query engine (typically supporting XPath and XQuery).

XML data warehouse: XML database that is specifically modeled (i.e., multidimensionally, with a star-like schema) to support XML decision-support and analytic queries.

Complex data: data that present several axes of complexity for analysis, e.g., data represented in various formats, diversely structured, from several sources, described through several points of view, and/or versioned.

Structural summary based-index: labeled graph structure that summarizes XML graph structural information.

Clustering: unsupervised machine learning method that consists in assigning a set of observations into subsets (clusters) so that observations in the same cluster are similar.

XML graph: data model representing the hierarchical nature of XML data. In a XML graph, nodes represent elements or attributes.